\documentclass[aps,twocolumn,pra,showpacs,superscriptaddress]{revtex4-1}
\usepackage{amsfonts}

\usepackage{graphicx,color}
\usepackage{epstopdf}
\usepackage{epsfig}
\usepackage[breaklinks=true]{hyperref}
\usepackage{graphicx}
\usepackage{mathrsfs}

\begin{document}

\title{How To Build A Quantum Coherent Observer}

\author{John E.~Gough}
\email{jug@aber.ac.uk}
   \affiliation{
	Aberystwyth University, SY23 3BZ, Wales, United Kingdom}

\author{Nina H. Amini}
\email{nina.amini@lss.supelec.fr}
      \affiliation{Laboratoire des Signaux et Syst\'{e}mes, CNRS, 91192 Gif sur Yvette, France}

\begin{abstract}
We give an explicit construction for a quantum observer coherently mimicking the dynamics
of a cavity mode system and without any disturbance of the system's dynamics. This gives the
exact analogue of the Luenberger observer used in controller design in engineering.
\end{abstract}

\maketitle

\affiliation{Institute for Mathematics and Physics, Aberystwyth University,
SY23 3BZ, Wales, United Kingdom}

\affiliation{Laboratoire des Signaux et Syst\'{e}mes, CNRS, 91192 Gif sur
Yvette, France}

\section{Introduction}
In this letter we wish to discuss an
engineering-oriented aspect of the quantum observers. In control theory, an
observer is a dynamical system capable of mimicking the dynamical state of a given
system (known as the plant): it should have the same class of state
variables and these variables should become close in the asymptotic
long-time limit. The concept was introduced by Luenberger \cite{Luenberger64}%
,\cite{Luenberger66} and plays an important role in controller design. 
In the quantum setting, an
observer receives information about the system. This can happen when the observer is making continuous
measurements on a quantum system and then deriving the
conditioned state of system using a quantum filter (quantum trajectories).
In many cases a quantum control problem can be broken down by a separation
principle \cite{BvH_sep} into a measurement stage
and an actuation stage. 

However, our interest is in quantum coherent observers. Here we mean a quantum
system that is coupled to the quantum (plant) system of interest and which
has capable of realizing a model of the plant's dynamics internally and
where this internal observer dynamics converges to that of the plant. The
concept was introduced in \cite{Zibo12} and developed as a conceptual device
for quantum design \cite{Zibo_16}-\cite{Pan16}. In these approaches, one
considers the observer embedded in a quantum feedback network with
connections to the plant system in such a way that there is a feedback loop
between the plant and observer - moreover the observer is then part of the
controller design problem. We wish to avoid this and have the plant
feedforward information to the observer - but not the other way round. In
this manner, there is no back-action of the observer on the plant - more
exactly, as no measurement need be involved, we mean that the plant's
dynamics is not modified by connection to the observer. Our construction,
given below, is based on the original Luenberger set-up and the quantization
is done by replacing the classical block design with a quantum feedback
network where the plant and observer are quantum linear systems and where
jump-off and summing junctions are replaced with beam-splitters.

\subsection{Classical Luenberger Observers}

\begin{figure}[htbp]
\centering
\includegraphics[width=0.3\textwidth]{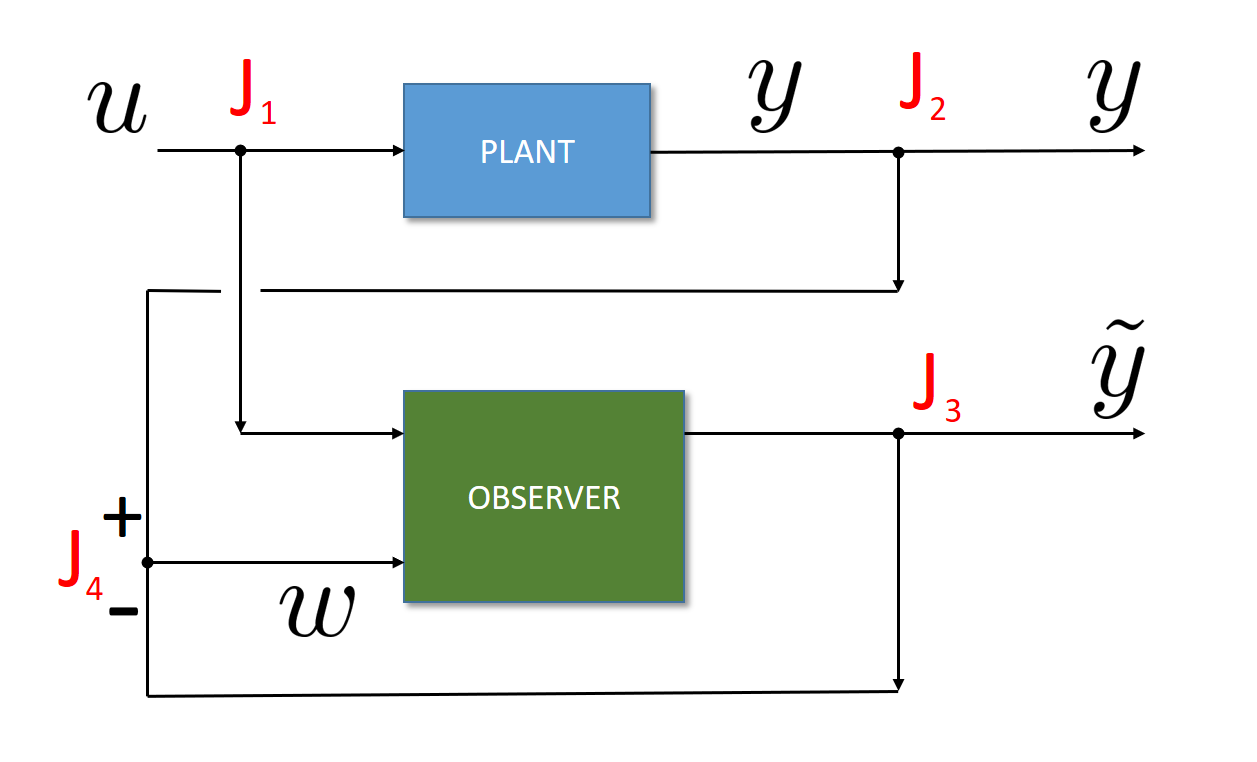}
\caption{A Luenberger observer connected to a plant system.}
\label{fig:Luenberger_classical}
\end{figure}

The set up for a classical Luenberger observer is sketched in Figure \ref
{fig:Luenberger_classical} which shows a plant system with input $u$ and
output $y$ connected to a second system which we term the observer.

The plant is taken to be a linear system with the input-state-output
equations 
\begin{eqnarray}
\dot{x} &=&Ax+Bu,  \nonumber \\
y &=&Cx.  \label{Eq:classical_ABC_plant}
\end{eqnarray}
What we wish to do is to have the observer track the state $x$ of the plant
system. The observer is also taken to be a linear system with state $\tilde{x%
}$, input $y$ and $w$ and output $\tilde{y}$ and input-state-output
equations 
\begin{eqnarray}
\dot{\tilde{x}} &=&A\tilde{x}+Bu+Lw,  \nonumber \\
\tilde{y} &=&C\tilde{x}.  \label{Eq:classical_ABC_observer}
\end{eqnarray}
Note that the same coefficients $A,B,C$ occur in both the plant and observer
models. The additional coefficient $L$ in the observer is called the \textit{%
Luenberger gain}.

In the set-up we require 4 junctions: $J_1 ,J_2, J_3$ are jump points where
we copy the signals $u$, $y$ and $\tilde{y}$ respectively; while $J_4$ is a
summing point where we subtract $\tilde{y}$ from $y$: $
w = y - \tilde{y} $.  

The \textit{error} between the plant state $x$ and the observer state $\tilde{x}$ is 
$e = x - \tilde{x} $, 
and we note that $w \equiv C \, e$.

Combining the equations (\ref{Eq:classical_ABC_plant}) and (\ref
{Eq:classical_ABC_observer}) we find 
\begin{eqnarray}
\dot{ e} &=& (A -LC) \, e,  \label{Eq:classical_error}
\end{eqnarray}
If $A-LC$ has strictly negative real part (Hurwitz in the multi-dimensional
case) then the error vanishes exponentially for long time, so we have
observer tracking the plant state for all initial conditions. More generally
we say that the plant is detectable if the pair $(A,C)$ has the property
that we may find a gain $L$ such that $A-LC$ is Hurwitz. In other words
detectability of a plant means that we may construct such an observer with
the error decreasing to zero.

We note that if we want verification of the convergence of the observer
state variable $\tilde{x}$ to the plant state variable $x$ then we can look
at the difference $y -\tilde{y}$ of the plant and observer outputs. This
equates with $w$ and, as we have seen, this is $Ce$. therefore if $w(t) \to
0 $ as $t \to \infty$ then we have confirmation that $e(t) \to 0$. (In the
multi-dimensional case we may have to make do with supporting evidence if $C$
is not full rank.)

\section{Quantum Linear Systems}
The inputs are modeled as quantum input process $b_{\text{%
in},j}$ and these satisfy singular commutation relations $[b_{\text{in}%
,j}(t) , b_{\text{in},k}(s)^\ast ] = \delta_{jk} \, \delta (t-s)$. 

Hudson-Parthasarathy developed a theory of quantum stochastic calculus
generalizing the It\={o} theory and an unitary evolution underlying the
above model will be given by the process $U\left( t\right) $ satisfying the
quantum stochastic differential equation  \cite{HP84}, \cite{GZ}
\begin{eqnarray}
dU\left( t\right) &=&\bigg\{ \sum_{k=1}^{n}L_{k}\otimes dB_{\text{in},k}^{\ast
}\left( t\right) -\sum_{k=1}^{n}L_{k}\otimes dB_{\text{in},k}\left( t\right)
\nonumber \\
&&-\big( \frac{1}{2}\sum_{k=1}^{n}L_{k}^{\ast }L_{k}+iH\big) \otimes
dt \bigg\} U\left( t\right)  \label{eq:HP_QSDE}
\end{eqnarray}
where $B_{\text{in},k}\left( t\right) $ is the annihilation process
(integral of $b_{\text{in},k}\left( t\right) $), and $H=H^{\ast }$, $%
L_{1},\cdots ,L_{n}$ are operators on the system space. The unitary
determines the system evolution according to $X\rightarrow U\left( t\right)
^{\ast }\left( X\otimes I\right) U\left( t\right) $ for each system operator 
$X$. It also yields the input-output relations as the outputs are determined
by $B_{\text{out},k}\left( t\right) =U\left( t\right) ^{\ast }\left(
I\otimes B_{\text{in},k}\left( t\right) \right) U\left( t\right) $.


A linear class is obtained by taking the system to consist of $m$
oscillators with mode operators $a_{1},\cdots ,a_{m}$ with Hamiltonian 
$H\equiv \sum_{\alpha =1}^{m}\sum_{\beta =1}^{m}\omega _{\alpha \beta
}a_{\alpha }^{\ast }a_{\beta }$,
and 
$L_{k}\equiv \sum_{\alpha =1}^{m}C_{k\alpha }^{-}a_{\alpha }+\sum_{\alpha
=1}^{m}C_{k\alpha }^{+}a_{\alpha }^{\ast }$.
The matrices $\Omega _{-}=\left[ \omega _{\alpha \beta }\right] \in \mathbb{C}%
^{m\times m}$ (Hermitean), and $C_{\pm }=\left[ C_{k\alpha }^{\pm }\right]
\in \mathbb{C}^{n\times m}$ then determine the model. For $R=\left[ R_{jk}%
\right] $ an array of operators, let us use the notation $R^{\ast },R^{\top
} $ and $R^{\#}$ for the adjoint $\left[ R_{kj}^{\ast }\right] $, the
transpose $\left[ R_{kj}\right] $ and the adjoint-transpose $\left[
R_{jk}^{\ast }\right] $ respectively. With column vectors $\mathbf{a}%
\left( t\right) =\left[ a_{\alpha }\left( t\right) \right] $, $\mathbf{b}_{%
\text{in}}\left( t\right) =\left[ b_{\text{in},k}\left( t\right) \right] $
and $\mathbf{b}_{\text{out}}\left( t\right) =\left[ b_{\text{out},k}\left(
t\right) \right] $ we have 
\begin{eqnarray}
\frac{d}{dt}\mathbf{a}\left( t\right) &=&A_{-}\mathbf{a}\left( t\right)
+A_{+}\mathbf{a}\left( t\right) ^{\#}  \nonumber \\
&& \qquad +B_{-}\mathbf{b}_{\text{in}}\left( t\right) +B_{+}\mathbf{b}_{%
\text{in}}\left( t\right) ^{\#},  \nonumber \\
\mathbf{b}_{\text{out}}\left( t\right) &=&C_{-}\mathbf{a}\left( t\right)
+C_{+}\mathbf{a}\left( t\right) ^{\#}+\mathbf{b}_{\text{in}}\left( t\right) ;
\label{eq:Gen_lin}
\end{eqnarray}
where we will have 
\begin{eqnarray}
A_{-} &=&-\frac{1}{2}C_{-}^{\ast }C_{-}+\frac{1}{2}C_{+}^{\top
}C_{+}^{\#}-i\Omega _{-},  \nonumber \\
A_{+} &=&-\frac{1}{2}C_{-}^{\ast }C_{+}+\frac{1}{2}C_{+}^{\top }C_{-}^{\#}, 
\nonumber \\
B_{-} &=&-C_{-}^{\ast }, \qquad
B_{+} =-C_{+}^{\top }.
\end{eqnarray}

\subsection{Quantum Plant Systems}

We will consider linear models for quantum plants where $C_{-}$ is a fixed
matrix $C$ and $C_{+}\equiv 0$. This leads to the simplified equations 
\begin{eqnarray}
\frac{d}{dt}\mathbf{a}\left( t\right) &=&A\mathbf{a}\left( t\right) +B%
\mathbf{b}_{\text{in}}\left( t\right) ,  \nonumber \\
\mathbf{b}_{\text{out}}\left( t\right) &=&C\mathbf{a}\left( t\right) +%
\mathbf{b}_{\text{in}}\left( t\right) ;
\end{eqnarray}
where $A\equiv -\frac{1}{2}C_{-}^{\ast }C_{-}-i\Omega _{-}$ and $B=-C^{\ast
} $. Note that if we wish to add classical fields $\mathbf{u}\left( t\right) 
$ as a driving term then we do this by making the translation 
\begin{eqnarray}
\mathbf{b}_{\text{in}}\left( t\right) \rightarrow \mathbf{b}_{\text{in}%
}\left( t\right) +\mathbf{u}\left( t\right) ,  \nonumber
\end{eqnarray}
in the above. In the special case of a single oscillator ($m=1$) we will have $H=\omega
a^{\ast }a$ and $L_{1}=\sqrt{\gamma }a$ so that this simplifies further to 
\begin{eqnarray}
\frac{d}{dt}a\left( t\right) &=&-\left( \frac{1}{2}\gamma +i\omega \right)
a\left( t\right) -\sqrt{\gamma }b_{\text{in}}\left( t\right) ,  \nonumber \\
b_{\text{out}}\left( t\right) &=&\sqrt{\gamma }a\left( t\right) +b_{\text{in}%
}\left( t\right) .
\label{eq:qplant}
\end{eqnarray}

We note that both the inputs and the outputs are quantum processes to which
we are adding a classical signal. The processes correspond to quantum
electromagnetic fields in the Markov regime. 

\subsection{A First Approach}
\begin{figure}[htbp]
\centering
\includegraphics[width=0.480\textwidth]{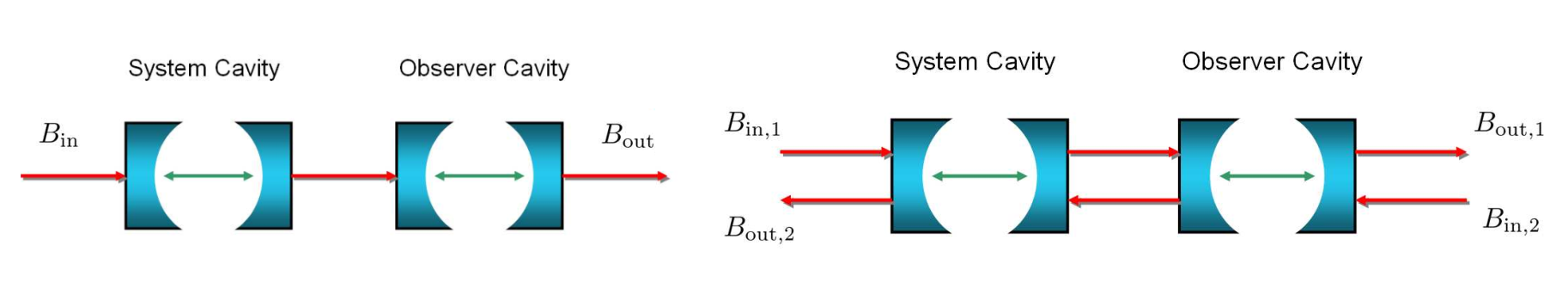}
\caption{(color online) Left: feedforward (one-way) from system cavity mode $a$ to observer cavity mode $\tilde{a}$;
Right: two-way interaction.}
\label{fig:OBS_oneway}
\end{figure}

As a first step, we consider the plant system (cavity mode $a$) and observer
(cavity mode $\tilde{a}$) as cascaded open systems
as depicted in Figure \ref{fig:OBS_oneway} (left).

With this feed-forward situation we have
\begin{eqnarray*}
\frac{d}{dt} \left[ 
\begin{array}{c}
a (t) \\ 
\tilde{a} (t)
\end{array}
\right] &=& A_1 \left[ 
\begin{array}{c}
a (t) \\ 
\tilde{a} (t)
\end{array}
\right] \,  - \sqrt{\gamma} \left[ 
\begin{array}{c}
1 \\ 
1
\end{array}
\right] \, b_{\mathrm{in}} (t) ,\\
b_{\mathrm{out}} (t) &=& b_{\mathrm{in}} (t) + \sqrt{\gamma} [ a (t)
+\tilde{a} (t) ] .
\end{eqnarray*}
where $A_1= \left[ 
\begin{array}{cc}
-\frac{1}{2} \gamma - i \omega & 0 \\ 
- \gamma & -\frac{1}{2} \gamma - i \omega
\end{array}
\right] $.
The combined system is stable as the matrix $A_1$ is Hurwitz: in particular it
has a double eigenvalue $-\frac{1}{2} \gamma - i \omega $. The stability
therefore implies that we have both oscillators damped to zero on average.

If however, we introduce a second channel running in the reverse direction, 
see Figure \ref{fig:OBS_oneway} (right), then we find
\begin{eqnarray*}
\frac{d}{dt} \left[ 
\begin{array}{c}
a (t) \\ 
\tilde{a} (t)
\end{array}
\right] &=& A_2 \left[ 
\begin{array}{c}
a (t) \\ 
\tilde{a} (t)
\end{array}
\right]    - \sqrt{\gamma} \left[ 
\begin{array}{c}
1 \\ 
1
\end{array}
\right] \, [b_{\mathrm{in},1} (t) + b_{\mathrm{in},2} (t) ] ,\\
\left[ 
\begin{array}{c}
b_{\mathrm{out}, 1} (t) \\ 
b_{\mathrm{out}, 2} (t)
\end{array}
\right] &=& \left[ 
\begin{array}{c}
b_{\mathrm{in}, 1} (t) \\ 
b_{\mathrm{in}, 2} (t)
\end{array}
\right]   + \sqrt{\gamma} [ a (t) +\tilde{a} (t) ] \left[ 
\begin{array}{c}
1 \\ 
1
\end{array}
\right]   .
\end{eqnarray*}
where now $
A_2= \left[ 
\begin{array}{cc}
-\frac{1}{2} \gamma - i \omega & -\frac{1}{2} \gamma \\ 
-\frac{1}{2} \gamma & -\frac{1}{2} \gamma - i \omega
\end{array}
\right]$.
The two-way cascade is now only marginally stable as the matrix $A_2$ now has acquired
a purely imaginary eigenvalue $-i \omega$, and has only the one eigenvalue $-%
\frac{1}{2} \gamma - i \omega $ with negative real part.

The modes have the explicit form 
\begin{eqnarray*}
a(t) &=&\frac{1}{2}e^{-i\omega t}[(1+e^{-\gamma t})a(0)+(1-e^{-\gamma
t}) \tilde{a} (0)]-\mathcal{A}_{t},  \nonumber \\
\tilde{a} (t) &=&-\frac{1}{2}e^{-i\omega t}[(1-e^{-\gamma t})a (0)-(1+e^{-\gamma
t})\tilde{a} (0)]-\mathcal{A}_{t}.  \nonumber \\
\end{eqnarray*}
where $
\mathcal{A}_t = \sqrt{\frac{\gamma}{2}} \int_0^t e^{- (\gamma +i \omega)
(t-s) } [ d B_{\mathrm{in},1} (s) + d B_{\mathrm{in},2} (s) ] $.

The mode $-\tilde{a}(t)$ is converging asymptotically to $a(t)$ in the sense that
the error $e(t) = a(t) +\tilde{a} (t)$ is given by
\[
e(t)= e^{- (\frac{1}{2} \gamma + i \omega )t}\, e(0) -2 \mathcal{A}_t .
\]
It is a first sight strange that the introduction of a second source of damping should result
in less damping, however, it is well-known in control theory that a network of systems that are
separately stable may itself be unstable. In particular, the quantum variable $a(t) - \tilde{a}(t)$ in the two-way network simply executes a harmonic motion. In fact, it determines a decoherence free subspace. A switching mechanism between the one-way and two-way set-up has recently been proposed
as a switch between the writing/read-out and storage configurations for a quantum memory scheme
\cite{NG15}.
It has also been used as an example of a design decoherence free subspace \cite{Pan16}.

Arguably, the two-way set-up is not the form we want. To begin with, we have altered the dynamics of
the plant by feedback from the observer which is something we wanted to avoid. We also have the
unwanted symmetry that the plant observes the observer as much as the observer observes the plant.
We will rectify this in the following constructions.

\section{Quantum Luenberger Observers}

\begin{figure}[htbp]
\centering
\includegraphics[width=0.450\textwidth]{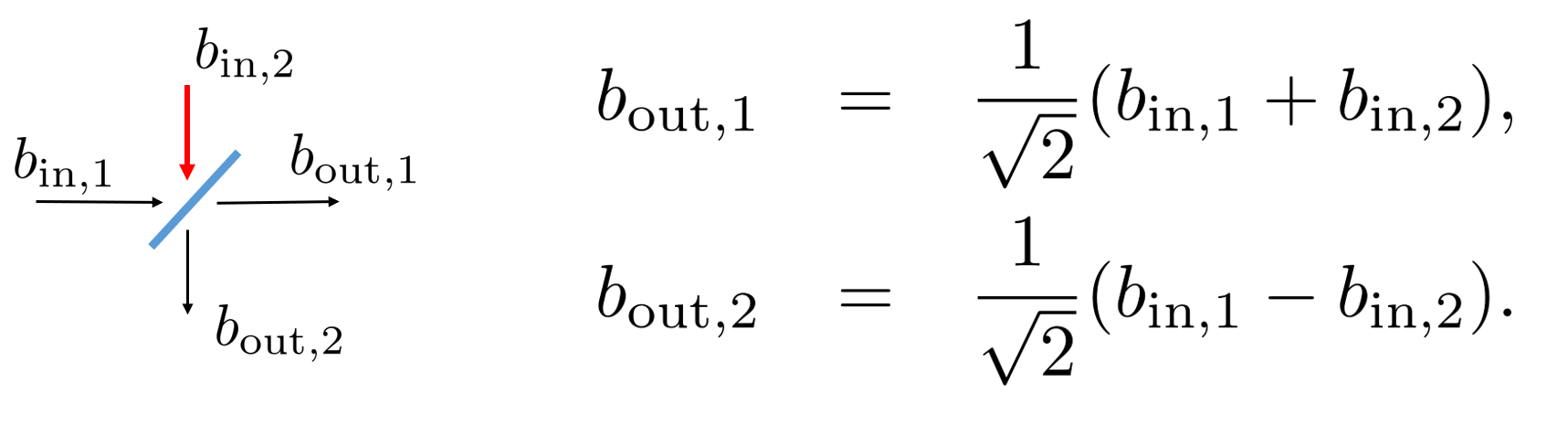}
\caption{(color online) 50-50 beam-splitter.}
\label{fig:Beamsplitter}
\end{figure}

We now consider the problem of construction a quantum version of the
Luenberger observer. Unfortunately, if we wished to
implement the classical Luenberger observer, as in Figure \ref
{fig:Luenberger_classical}, then we run into the problem that we cannot
clone quantum information. Instead, the junctions $J_1, J_2,J_3$ have to be
replaced by beam-splitters. We do the same for $J_4$ in order to subtract
the two quantum processes. In each case, we will replace the junction by a
50-50 beam-splitter performing the transformation (see Figure \ref
{fig:Beamsplitter} ).

Note that this may involve introducing additional noises (indicated in red).

\subsection{Quantum Luenberger Observers}

\begin{figure}[tbph]
\centering
\includegraphics[width=0.3\textwidth]{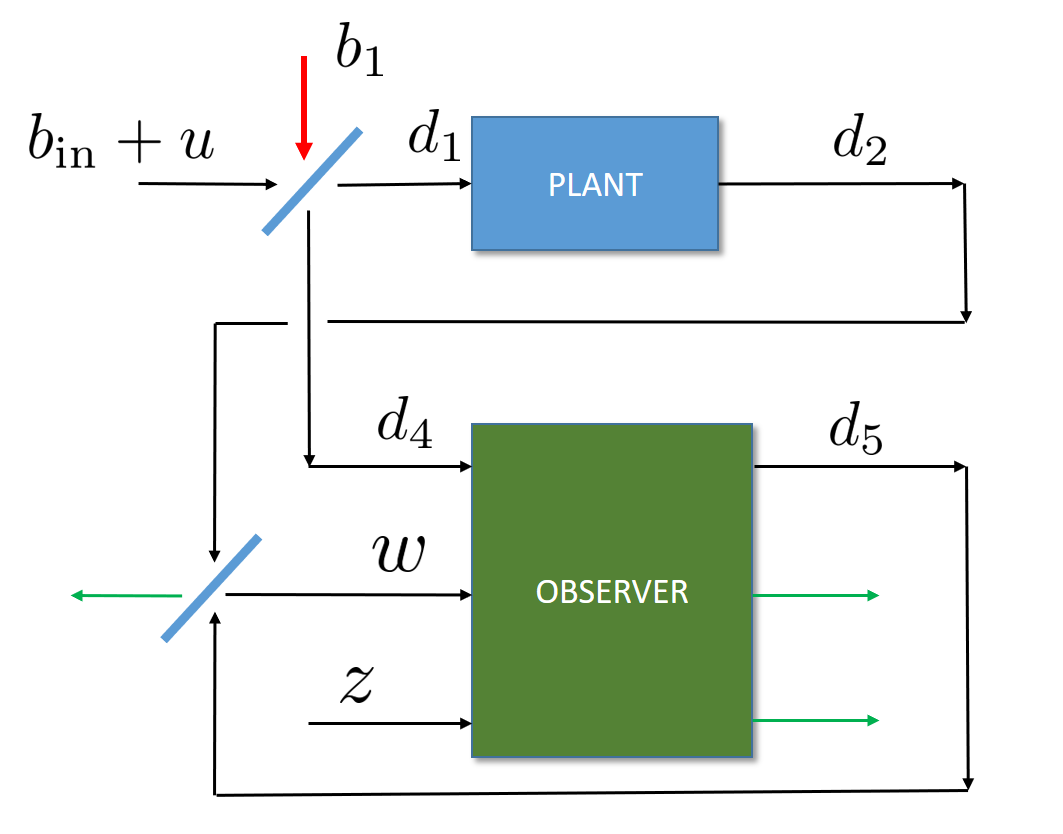}
\caption{(color online) A quantum plant and observer}
\label{fig:Luenberger_quantum_unverified}
\end{figure}

We first set about replacing junctions $J_{1}$ and $J_{4}$ with
beam-splitters. To keep things simple, we ignore junctions $J_{2}$ and $J_{3}
$ though this means that we cannot now pass the relevant outputs of the
plant and observer to the outside world as before. The situation is depicted
in Figure \ref{fig:Luenberger_quantum_unverified}. Note that we need to
introduce a new independent quantum input process $b_{1}$ (highlighted in
red). The plant will be a cavity mode $a$ with frequency $\omega $ and a
single input leading to a damping rate $\gamma $, (so $A=-\frac{1}{2}\gamma
-i\omega ,B=-\sqrt{\gamma },C=\sqrt{\gamma }$).

For the observer, we need a minimum of two inputs $b_{\text{in},1}\equiv
d_{4}$ and $b_{\text{in},2}\equiv w$, see Figure \ref
{fig:Luenberger_quantum_unverified}, with coupling operators $L_{1}=\sqrt{%
\gamma }\tilde{a}$ and $L_{2}=\sqrt{\gamma _{L}}\tilde{a}$. Here $L=\sqrt{%
\gamma _{L}}$ is the \textit{quantum Luenberger gain coefficient}. However,
on its own, this would lead to $A^{\text{obs}}=-\frac{1}{2}(\gamma +\gamma
_{L})-i\omega $ so we have introduced more damping into the observer than in
the plant leading to a different $A$ matrix. The choice of $L_{1}$ is
constrained to ensure that the $B$ and $C$ coefficients of the plant and
observer match up, so we cannot adjust this. 

The only other option is to
introduce a third input $b_{\text{in},3}$ whose role is to supply energy to
compensate for $b_{\text{in},2}$: this is achieved by taking the associate
coupling to be $L_{3}=\sqrt{\gamma _{L}}\tilde{a}^{\ast }$. The observer is
then described by
\begin{eqnarray}
\frac{d}{dt}\tilde{a}\left( t\right)  &=&-\left( \frac{1}{2}\gamma +i\omega
\right) \tilde{a}\left( t\right) -\sqrt{\gamma }b_{\text{in},1}\left(
t\right) \nonumber \\
&& -\sqrt{\gamma _{L}}b_{\text{in},2}\left( t\right) -\sqrt{\gamma _{L}%
}b_{\text{in},3}\left( t\right) ^{\ast } 
\label{eq:observer_QDuV}
\end{eqnarray}
with $b_{\text{out},1}\left( t\right)  =\sqrt{\gamma }\tilde{a}\left( t\right)
+b_{\text{in},1}\left( t\right) $, 
$b_{\text{out},2}\left( t\right)  =\sqrt{\gamma _{L}}\tilde{a}\left(
t\right) +b_{\text{in},2}\left( t\right)$, \and
$b_{\text{out},3}\left( t\right)  =\sqrt{\gamma _{L}}\tilde{a}\left(
t\right) ^{\ast }+b_{\text{in},3}\left( t\right) $.

The outputs $b_{\text{out},2}\left( t\right) $ and $b_{\text{out},3}\left(
t\right) $ are ignored, however, we see comparing (\ref{eq:qplant}) and (\ref{eq:observer_QDuV}) that the $A$
coefficients are equal, as are the $B$ and $C$ coefficients or $b_{\text{in}%
}\left( t\right) $ and $b_{\text{in},1}\left( t\right) $.

The relevant equations, with reference to Figure \ref{fig:Luenberger_quantum_unverified}, are
\begin{eqnarray*}
&&( J_1 \, \text{Beam-splitter})  \qquad \left\{ 
\begin{array}{c}
d_{1}=\frac{1}{\sqrt{2}}(b_{\text{in}}+u)+\frac{1}{\sqrt{2}}b_{1}, \\ 
d_{4}=\frac{1}{\sqrt{2}}(b_{\text{in}}+u)-\frac{1}{\sqrt{2}}b_{1};
\end{array}
\right.  \\
&&(J_4 \, \text{Beam-splitter}) \qquad w=\frac{1}{\sqrt{2}}d_{2}-\frac{1}{\sqrt{2}}d_{5}; \\
&&(\text{Plant})\quad \left\{ 
\begin{array}{c}
\dot{a}=Aa+Bd_{1} \\ 
d_{2}=Ca+d_{1};
\end{array}
\right.  \\
&&(\text{Observer})\quad \left\{ 
\begin{array}{c}
\dot{a}=A\tilde{a}+Bd_{4}-Lw, \\ 
d_{5}=C\tilde{a}+d_{4}.
\end{array}
\right. 
\end{eqnarray*}
We now introduce the error operator defined to be $e=a-\tilde{a}$ and after
some algebra we find $\dot{e} =\big(A-\frac{1}{\sqrt{2}}LC\big)e\left( t\right) -\sqrt{2}\left(
C+L\right) b_{1}\left( t\right)$, or
\begin{eqnarray*}
\dot{e} 
\equiv \left( -\frac{1}{2}\gamma -\sqrt{\frac{\gamma \gamma _{L}}{2}}%
+i\omega \right) e\left( t\right) -\sqrt{2}\left( \sqrt{\gamma }+\sqrt{%
\gamma _{L}}\right) b_{1}\left( t\right) .
\end{eqnarray*}

\subsection{Verifiable Quantum Luenberger Observers}
\begin{figure}[htbp]
\centering
\includegraphics[width=0.35\textwidth]{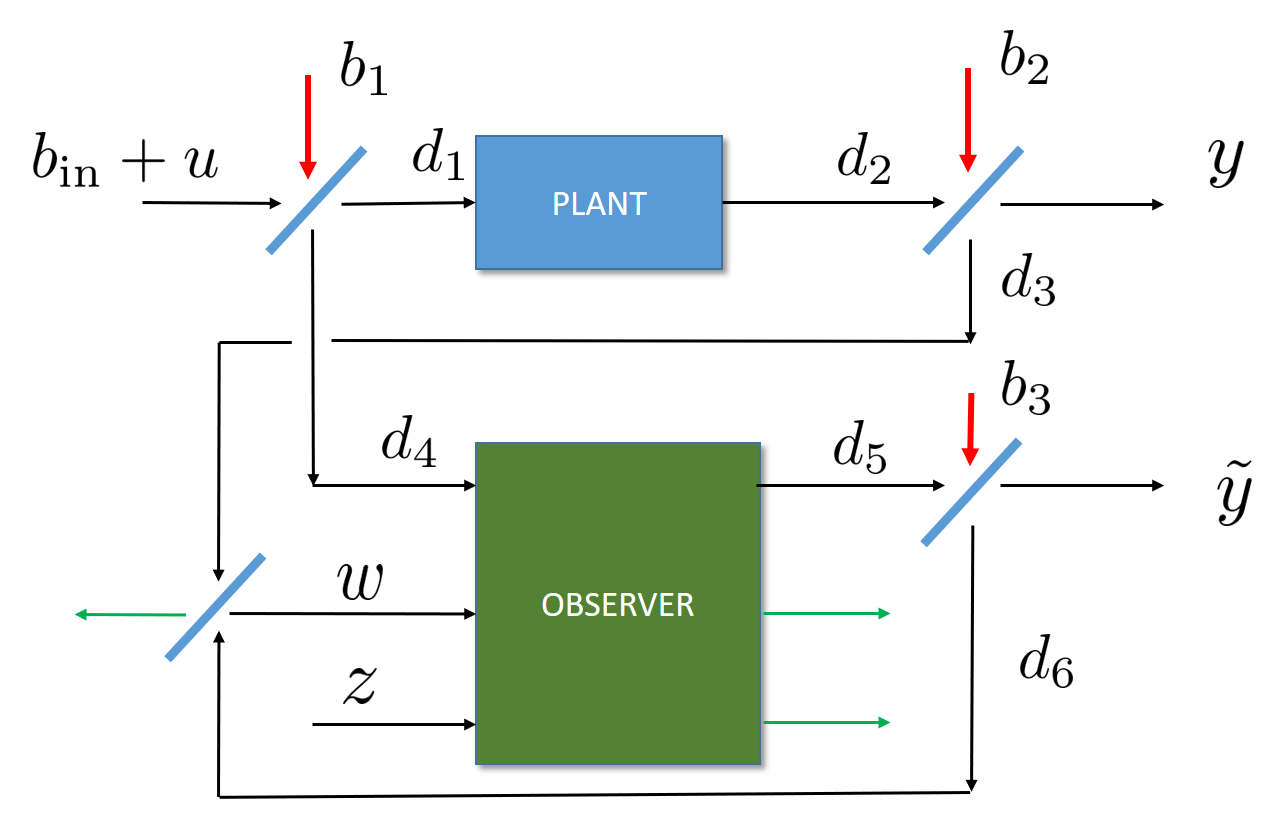}
\caption{(color online) Verifiable quantum observer}
\label{fig:Luenberger_quantum}
\end{figure}

If we wish to verify the convergence then we may consider the setup in Figure
\ref{fig:Luenberger_quantum}. This time, $
\dot{e} (t)=\big( A-\frac{1}{\sqrt{2}}LC \big) e\left( t\right) -\sqrt{2}\left(
C+ \frac{1}{2} L\right) b_{1} ( t )  -\frac{1}{2} Lb_2 (t) + \frac{1}{2} L b_3 (t) -L z^\ast(t)$.

We may send the outputs $y$ and $\tilde{y}$ into a 50-50 beam-splitter and measure the output
\begin{eqnarray}
\frac{y(t) -\tilde{y}(t)}{\sqrt{2}} =
\frac{1}{\sqrt{2}} C \, e(t) + b_1 (t) + \frac{1}{\sqrt{2}} b_2 (t) -\frac{1}{\sqrt{2}} b_3 (t) .\nonumber \\
\end{eqnarray}
Note that this output is likewise unaffected by any input disturbance $u$ and on average decays to zero as
$\langle e(t) \rangle \to 0$ for large time.

\section{Conclusion}
We have given an explicit construction for a coherent quantum observer which observes a
quantum system without altering its dynamics. In common with the classical Luenberger
observer, any disturbance $u$ carried into the plant system does not get transferred to the
observer. All inputs act passively on the cavity modes, except the input $z$ which supplies an active element
in the observer to compensate the additional damping and ensure that the appropriate observer $A-B-C$ terms agree
with those of the plant system.

The observer may be physically realized, however it is also of conceptual value in observer-based design of quantum coherent
feedback controllers.

\end{document}